\documentclass{revtex4}
\usepackage{graphicx}
\usepackage{dcolumn}
\usepackage{amsmath}
\usepackage{amssymb}
\usepackage{amsfonts}
\usepackage{bm}
\usepackage{color}
\newcommand{\be}{\begin{equation}}
\newcommand{\ee}{\end{equation}}
\newcommand{\ba}{\begin{eqnarray}}
\newcommand{\ea}{\end{eqnarray}}

\begin{document}
\title{Designing drones by combining finite element and atomistic simulations:
a didactic approach}
	
\author{Marcello Raffaele$^{\#}$}
\author{Maria Teresa Caccamo$^{\dag}$}
\author{Giuseppe Castorina$^{\dag}$}
\author{Stefania Lanza$^{\dag}$}
\author{Salvatore Magaz\`u$^{\dag}$}
\author{Gianmarco Muna\`o$^{\dag}$}
\author{Giovanni Randazzo$^{\dag}$}
\thanks{Corresponding author, email: {\tt grandazzo@unime.it}}

\affiliation{
$^{\#}$Dipartimento di Ingegneria,
Universit\`a degli Studi di Messina,
Viale F.~Stagno d'Alcontres 31, 98158 Messina, Italy. \\
$^{\dag}$Dipartimento di Scienze Matematiche e Informatiche, Scienze Fisiche
e Scienze della Terra,
Universit\`a degli Studi di Messina,
Viale F.~Stagno d'Alcontres 31, 98158 Messina, Italy.}

\maketitle

\section*{Abstract}
A didactic multiscale approach for drone modeling is proposed. Specifically,
we investigate the drone structure at both macroscopic and microscopic scales,
by making use of finite element and atomistic simulations, respectively.
The structural analysis is performed with the aim to equip the drone with
specific sensors and measuring instruments capable to detect the existence
of volcanic ash, SO$_2$, CO$_2$ and other pollutants in the atmosphere after a 
vulcanic eruption. We show that, by modeling the tubular structure of the 
drone with a sandwich constituted by a a polystyrene core, carbon fiber 
skins and epoxy matrix, a weight saving of 7 grams for each drone arm can be
obtained, in comparison to the standard commercial drones, although
a slight worsening of the mechanical performances is observed. 
In addition, the molecular structure of the polystyrene chains 
has been investigated by using atomistic Molecular Dynamics simulations, 
providing further information on the local structure of the polymer chains.
Additional improvements of the weight saving could be obtained 
by means of the the topological optimization techniques on the body of the 
drone and on the supports for landing.

\section{Introduction}
The mechanical design of lightweight structures allows to obtain, through a 
topological and material optimization, a lighter and more efficient 
structure~\cite{Sigmund:13}, capable of maintaining the same mechanical 
performances as the native material, and in some cases improving them. 
In the course of the last fifty years, composite materials such as carbon 
fiber or kevlar have supplanted the use of metals as aluminum in the creation 
of cockpits in various competitive sports as well as in the automotive 
and offshore sectors~\cite{Cucinotta:20}. 
Over the past decade, many steps forward have been made 
in additive manufacturing 
(AM)~\cite{Gardan:16} and thanks to topological optimization 
techniques, it is possible to produced lightweight structures 
quickly and efficiently~\cite{Cucinotta:19}.
A widespread application of such a technology is constituted by drones, which
can be used in a large variety of context, from simple entertainment to the
monitoring of meteorological conditions and atmospheric pollutants.
For instance, it is possible to equip drones with proper sensors capable 
of measuring the quantity of polluting gases in the upper atmosphere.
One of the building blocks constituting drones that requires a great care 
during the assembly phase is the frame. Indeed, the latter must be capable to
withstand the large amount of stress to which the drone is usually subjected
during the flight phase. The frame of a drone typically consists of a support
constitued by plastic material, aluminum and carbon.
The behavior of plastic material, in particular, is crucial in order to
provide the necessary flexibility to the frame; at the same time, this 
material must be strong enough to avoid breakages and also has to act as a
thermic and electrical insulator. \\
In the present work we didactically explain how is possible to model a drone
with the desired properties by making use of the finite elements simulations.
The proposed study is motivated by the possibility to use such a drone to
measure the amount of volcanic ash, SO$_2$ and CO$_2$ released in the atmosphere
during a volcanic eruption. In addition, the drone can also be used as an
efficient tool to provide early warning on the volcanic activity.
The possibility to perform these measurements 
would constitute a strong validation of numerical models, such as the 
Weather Research and Forecasting model coupled with chemistry 
(WRF-Chem)~\cite{Grell:05,Stuefer:13},
specifically developed to monitor the existence of pollutants in the 
atmosphere. The WRF-Chem package is the generalization of the meteorological
WRF model~\cite{Skamarock:08,Powers:17}, suited to investigate, among others, 
convective motions~\cite{Castorina:19}, 
nucleation processes~\cite{Restuccia:18} 
and heavy rainfall events~\cite{MTCaccamo:17}; further atmospheric phenomena can
be modeled by using theoretical approaches, such as those based on adiabatic
expansion and compression~\cite{MTCaccamo:19,Castorina:18}.\\
In addition, we also show how to investigate the microscopic behavior
of the plastic materials constituting the drone frame by using Molecular
Dynamics (MD) simulations which, along with experimental techniques like
X-ray scattering, has been largely used to investigate the fluid structure
of polymers and block copolymers~\cite{Lombardo:19}. 
Therefore, upon combining these two different methods,
we propose a simple approach to a multiscale modeling of drones.
Specifically, the MD-based investigation is focused on 
the polystyrene, 
a remarkable example of plastic material whose properties can be
succesfully detailed by using computer simulations. 
In this context, the 
MD technique has been largely adopted to perform simulation studies
of both polystyrene melts~\cite{Harmandaris:11,Denicola:16} 
and composite materials obtained by filling the
polymer matrix with inorganic 
particles~\cite{Muller-Plathe:12,Ndoro:11,Munao:18a,Munao:19} or 
nanotubes~\cite{Zhao:16,Donati:20}. 
The employment of the MD techniques allows to gain
knowledge into the structural and dynamical properties of polymer chains, 
which, in turn, determine the final properties of the material. In particular,
it is possible to estimate the chain relaxation, {\it i.e.} the time required
by a polymer chain to assume its equilibrium configuration, which strictly
depends on temperature, density and interactions with other chains and with
the solvent. In addition, the computation of the end-to-end distance,
gyration radius, chain orientation and density profiles allows to obtain
a general picture of the miscroscopic properties of the 
material~\cite{Munao:18a,Sparnacci:20}. \\
As far as the finite elements simulations are concerned,
the case study investigated in the present work 
is that of a commercial drone model DJI S900 for which we show 
how to make the supporting structure lighter, at the same time leaving the 
mechanical characteristics unchanged. This will allow for a greater battery 
autonomy which translates into longer flight time.
The finite element software used in this work is the Siemens Nx Nastran. \\
The didactic approach proposed in this work, based on the
``learning by doing'' method, is presented in the next section. 
The design of the drone structure is described in Section III, while
details on the computational approaches
suited to investigate the structural properties of
polystyrene chains are given in Section IV. The finite element simulations 
are presented and discussed in Section V, and finally conclusions 
follow in the last section.

\section{Didactic methodology: learning by doing}
The didactic approach proposed in the present manuscript can be fruitfully
addressed through the ``learning by doing'' method. The learner is not a 
simple, passive, observer, but actively participates to the realization of the
project target, at the same time learning the mechanisms behind it. 
Specifically, he will have to start designing the drone structure by following
the prescriptions of finite element simulations. Also, he will have the 
possibility to learn the basis of Molecular Dynamics simulations. The
required prerequisites are a basic knowledge of atomic and molecular 
chemistry, along with the knowledge of mechanics and statics of rigid bodies.
Starting from these basis, the learner will have the possibility to design the
system to be investigated; for instance, as detailed in Section IV, he will
have to built a simulation box containing a certain number of polystyrene 
chains. Starting from different possible configurations, he could decide,
for instance, to 
put many polymer chains of low molecular weight or rather few chains with high 
molecular weight: even if the total number of atoms
is the same, the simulations could provide different results. In this way
the learner will be able to infer which is the best compromise between
speed calculation and model accuracy. A good combination of these two aspects
is reported  in Section IV. A similar approach holds for the finite element
simulations, where the learner, upon modeling the tubular structure of the 
drone with a polystyrene matrix, will obtain a weight saving at the expense
of mechanical properties. 
According to this didactic approach, the main target is not to memorize the
procedure, bur rather to understand the whole mechanism, enabling the learner
to face new problems.

\section{The drone structure}
%%%%%%%%%%%%%%%%%%%%%%%%%%%%%%%%%%%%%%%%%%%%%%%%%%%%%%%%%%%%%%%%%
\begin{figure}
\begin{center}
\includegraphics[width=8.0cm,angle=0]{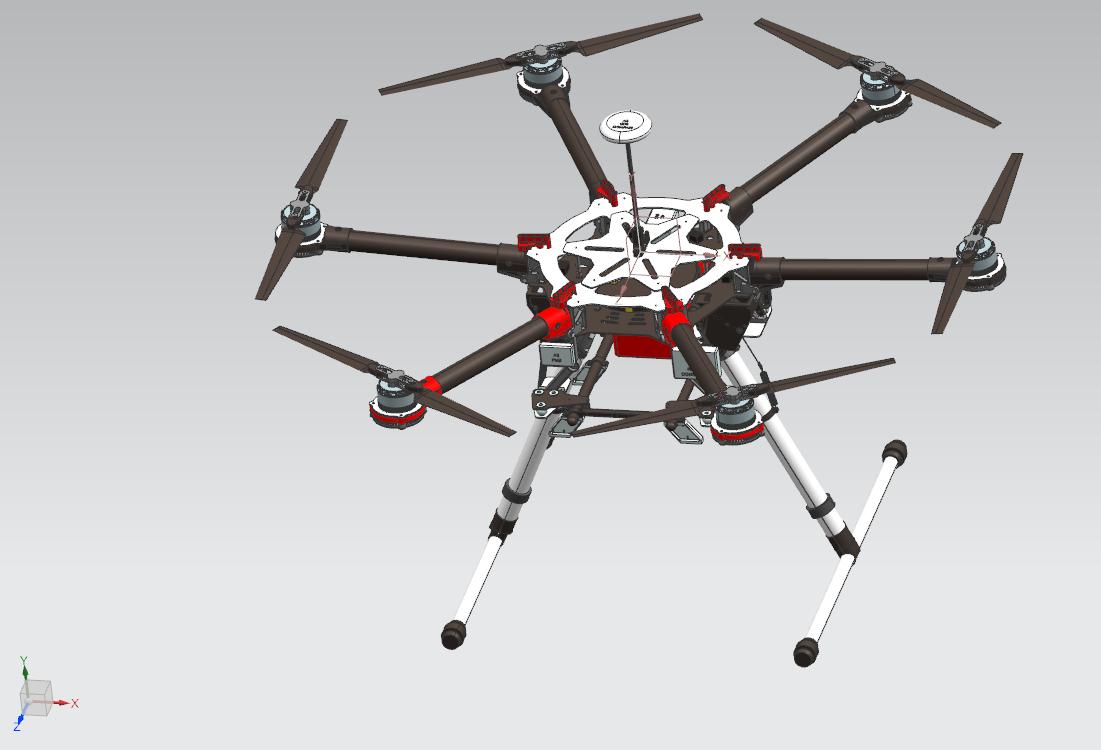} 
\caption{Reverse engineering of the drone DJI S900, investigated in this 
work.}\label{fig:drone1}
\end{center}
\end{figure}
\begin{figure}
\begin{center}
\includegraphics[width=8.0cm,angle=0]{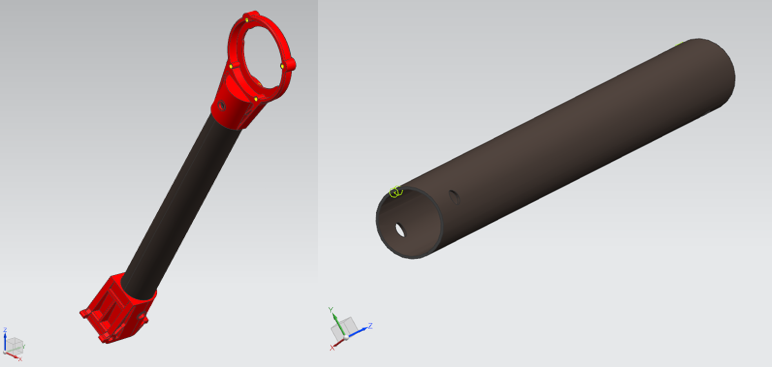} 
\caption{CAD of the support arm of one of the rotors.}\label{fig:drone2}
\end{center}
\end{figure}
\begin{figure}
\begin{center}
\begin{tabular}{cc}
\includegraphics[width=8.0cm,angle=0]{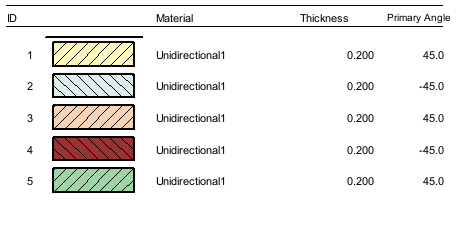} \\
\includegraphics[width=8.0cm,angle=0]{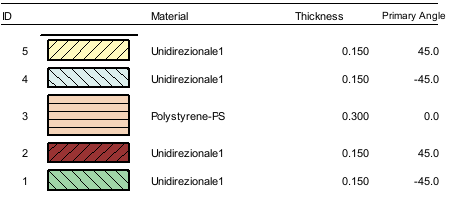} 
\end{tabular}
\caption{Structure of the laminate (top) and the sandwich (bottom).}	
\label{fig:drone34}
\end{center}
\end{figure}
\begin{figure}
\begin{center}
\includegraphics[width=8.0cm,angle=0]{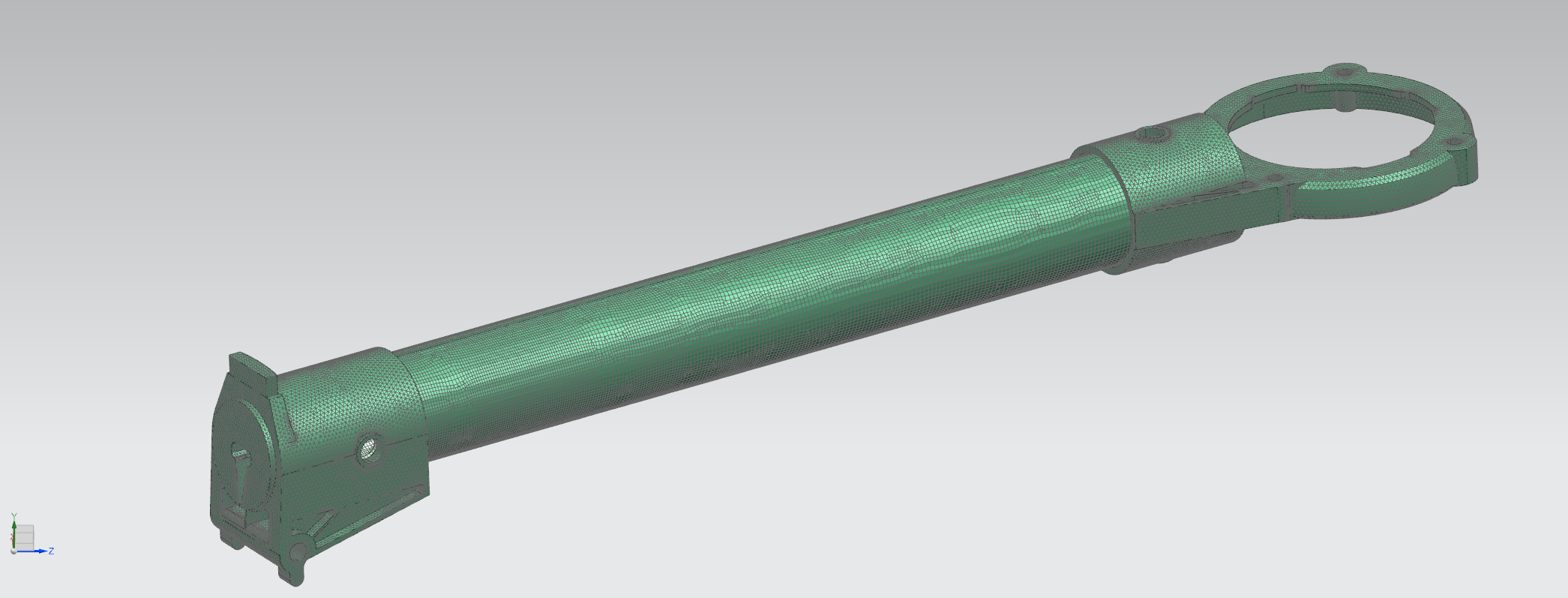} 
\caption{Illustration of the body mesh.}\label{fig:drone5}
\end{center}
\end{figure}
\begin{figure}
\begin{center}
\begin{tabular}{cc}
\includegraphics[width=7.0cm,angle=0]{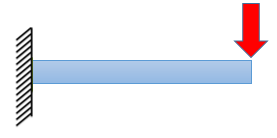} 
\includegraphics[width=7.0cm,angle=0]{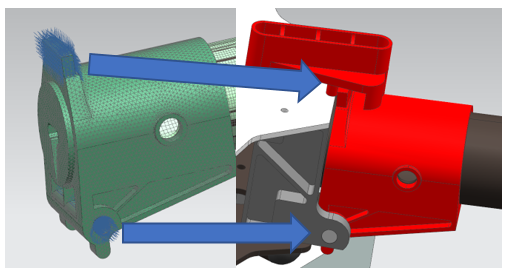} 
\end{tabular}
\caption{Left: schematic representation of the equivalent mechanical system.
Right: sketch of the boundary conditions and the joint of the drone arm.}
\label{fig:drone6}
\end{center}
\end{figure}

%%%%%%%%%%%%%%%%%%%%%%%%%%%%%%%%%%%%%%%%%%%%%%%%%%%%%%%%%%%%%%%%%
In the present work we simulate the realization of a drone after 
the replacement of 
the laminate of the supporting structure of the rotors (made up of only carbon 
fiber and epoxy resin) with a sandwich consisting of a polystyrene core 
(center) and a skin made of carbon fiber and epoxy resin; the two laminates 
will be compared both as regards the weight and the mechanical performances 
obtained. The use of polystyrene has already been studied either as the core 
of a sandwich or doped with other elements such as carbon 
nanotubes~\cite{Qian:00,Teodorescu:08,Potter:17}.
The drone considered in this work is the marketed DJI S900, a 
hexacopter with six arms that support the six rotors. The main 
features include a maximum motor power of 500 watts, a total weight (including
batteries) of 3.3 Kg and a takeoff weight of $\approx$ 8 Kg.
The tubular laminate of the drone arm (with a thickness of 1 mm) is realized 
in carbon fiber; the rotor fixing and arm anchoring to the body of the drone is
made in aluminum; the drone body for electronic control housing is constituted 
by plastic materials. The reverse engineering of the drone is reported in
Fig.~\ref{fig:drone1}.
The aim of the present study is to analyze the drone performaces upon 
replacing the epoxy matrix of the rotor            
bearing structure with a polystyrene matrix. 
The rotor bearing structure is formed by a tube with a thickness of 1 mm.
In Fig.~\ref{fig:drone2} it is possible to observe 
the Computer-Aided Design (CAD)
of one of the arms studied in 
this work: the red components are made in aluminum, while the tubular part is a 
laminate in carbon fiber.
%%%%%%%%%%%%%%%%%%%%%%%%%%%%%%%%%%%%%%%%%%%%%%%%%%%%%%%%%%%%%%%%%
%%%%%%%%%%%%%%%%%%%%%%%%%%%%%%%%%%%%%%%%%%%%%%%%%%%%%%%%%%%%%%%%%
The mechanical characteristics of the materials used for simulating the drone
realization are reported in Tab.~\ref{tab:drone-tab1}.
In particular, we show the values of densities $\rho$ (in g/cm$^3$), 
Young modulus E (in GPa), Poisson 
coefficient $\mu$ and yield stress $\sigma_{\mu}$ (in MPa) 
of fiber carbon, epoxy, polystyrene and
aluminum. The Poisson coefficient is defined as the ratio between transverse 
and longitudinal deformation, therefore is adimensional.
%%%%%%%%%%%%%%%%%%%%%%%%%%%%%%%%%%%%%%%%%%%%%%%%%%%%%%%%%%%%%%%%%
\begin{table}
\begin{center}
\caption{Mechanical properties of the materials used for simulating the drone.
Specifically, $\rho$ indicates the density, E the Young modulus, $\mu$  the 
Poisson coefficient and $\sigma_{\mu}$ the yield stress.}\label{tab:drone-tab1}
\begin{tabular}{cccccc}
\hline
& & $\rho$ [g/cm$^3$]  & E [GPa] & $\mu$ & $\sigma_{\mu}$[MPa] \\
\hline
& Fiber carbon & 1.8 & 240 & 0.2 & 4500 \\
& Epoxy & 1.3 & 3 & 0.37 & 27 \\
& Polystyrene & 1.01 & 3.5 & 0.45 & 30 \\
& Aluminum 6061 & 2.71 & 69 & 0.33 & 276 \\
\hline
\end{tabular}
\end{center}
\end{table}
%%%%%%%%%%%%%%%%%%%%%%%%%%%%%%%%%%%%%%%%%%%%%%%%%%%%%%%%%%%%%%%%%
The structure of the laminate (top panel of Fig.~\ref{fig:drone34}) consists of 
five layers with the 
orientation of the fibers at +/- $45^{\circ}$ and with a thickness of 0.2 mm 
for each layer.
%%%%%%%%%%%%%%%%%%%%%%%%%%%%%%%%%%%%%%%%%%%%%%%%%%%%%%%%%%%%%%%%%
%%%%%%%%%%%%%%%%%%%%%%%%%%%%%%%%%%%%%%%%%%%%%%%%%%%%%%%%%%%%%%%%%
In the bottom panel of Fig.~\ref{fig:drone34} it is shown the lamination 
plane of the sandwich. This is formed by a central core of polystyrene with 
a thickness of 0.3 mm and by two leathers placed both in the upper and lower 
part with fibers oriented to +/- 45$^\circ$ and thickness 0.15 mm.
The NX Nastran software has a practical built-in tool that allows  
to create the laminate. We start by defining the datasheets of the 
materials constituting the composite, followed by the matrix, the 
reinforcement fiber and the volumetric fraction, defined as the ratio between 
the volume of the fibers (or matrix) and the total volume of the 
composite. In the 
laminate investigated in the present work the volumetric fraction of the 
matrix has been set to $V_m = 0.4$ and for the fibers to $V_f = 0.6$.
After defining the materials used, we move on to the mesh phase, whose main 
feature is to discretize the continuous body in a grid made up of elements 
of various shapes (triangles, quadrilaterals for 2D domains or tetrahedrals 
for 3D mesh) connected together through points called nodes. This is done 
because in the calculation of the mechanical performance of a body, 
the calculator does nothing but
apply the laws of classical mechanics ($\it i.e.$ differential equations) 
to nodes. This procedure is necessary, since, upon applying them to 
a continuous body, we would obtain infinite results; conversely, by properly
discretizing the matter with a mesh, the mechanical behavior of the body can 
be approximated as close as possible to the real one, obtaining an advantage 
in the calculation speed. The denser the mesh (therefore the elements between 
the nodes are very small) the more precise the calculation will be at the 
expense of computational resources and computation time. As a consequence, 
a balance must be found between the mesh size and the computation time. 
In this work, the body was discretized with elements with a tetrahedral shape 
for the aluminum parts, while for the laminate quad elements were used and 
a size equal to 1 mm was set (see Fig.~\ref{fig:drone5}).
%%%%%%%%%%%%%%%%%%%%%%%%%%%%%%%%%%%%%%%%%%%%%%%%%%%%%%%%%%%%%%%%%
%%%%%%%%%%%%%%%%%%%%%%%%%%%%%%%%%%%%%%%%%%%%%%%%%%%%%%%%%%%%%%%%%
The boundary conditions of the system are two: the joint that holds the arm 
in the right position and the rotor that generates the lift force for pushing 
the drone upwards. The equivalent system is nothing more than a beam stuck 
on one side while the force is applied on the opposite side,
as schematicaly shown in the left panel of Fig.~\ref{fig:drone6}.
%%%%%%%%%%%%%%%%%%%%%%%%%%%%%%%%%%%%%%%%%%%%%%%%%%%%%%%%%%%%%%%%%
%%%%%%%%%%%%%%%%%%%%%%%%%%%%%%%%%%%%%%%%%%%%%%%%%%%%%%%%%%%%%%%%%
%%%%%%%%%%%%%%%%%%%%%%%%%%%%%%%%%%%%%%%%%%%%%%%%%%%%%%%%%%%%%%%%%
\begin{figure}
\begin{center}
\begin{tabular}{cc}
\includegraphics[width=7.0cm,angle=0]{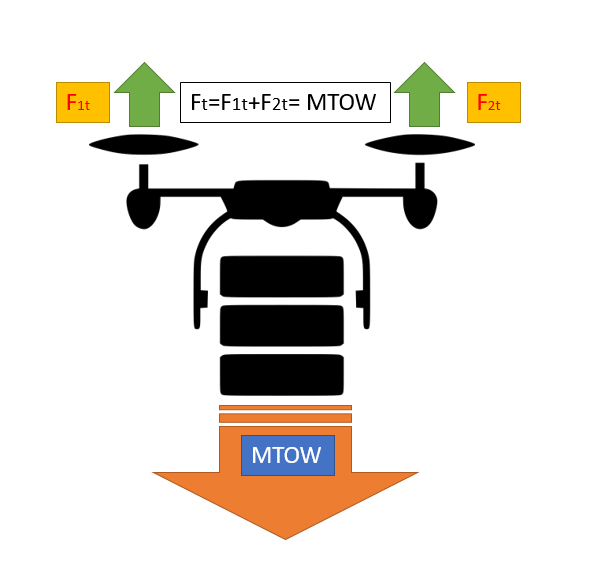} 
\includegraphics[width=7.0cm,angle=0]{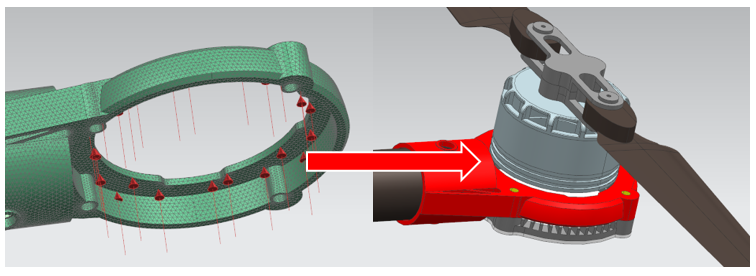} 
\end{tabular}
\caption{Left: schematic representation of the drone boost, 
necessary to overcome the maximum takeoff weight (MTOW). Right: boundary
conditions in the loading phase.}
\label{fig:drone8}
\end{center}
\end{figure}
%%%%%%%%%%%%%%%%%%%%%%%%%%%%%%%%%%%%%%%%%%%%%%%%%%%%%%%%%%%%%%%%%
The joint is shown in the right panel of Fig.~\ref{fig:drone6} 
and the drone arm 
remains in position thanks to a bottom hinge and a top stop colored in blue; 
this type of constraint blocks each of the six available degrees of freedom 
for each body.
As regards the boost exerted by the six rotors, 
an important datum is given by the maximum takeoff weight (MTOW), which 
indicates the maximum weight at which the drone is able to take off.
This value depends on the limits of the structure or on the limits of 
the motors: therefore in our study we have considered the most stressed 
condition, corresponding to a weight of 8 kg. In order to sustain the flight, the drone must have rotors capable of generating a boost at least equal to or 
greater than the weight transported, as shown in the left panel of 
Fig.~\ref{fig:drone8}. 
Given the maximum takeoff weight, the corresponding force that must be 
overcame is therefore:
\begin{equation}
F_t={\rm MTOW} \cdot g = 8 {\rm kg} \cdot 9.81 {\rm m/s^2} = 78.5 N
\end{equation}
where $g$ is the gravity acceleration.
This force value must be divided by the number of rotors, therefore for each 
arm that supports the rotor it must exist a force equal to 13 N 
(see right panel of Fig.~\ref{fig:drone8}); 
this in turn will stress the arm with a bending moment.
%%%%%%%%%%%%%%%%%%%%%%%%%%%%%%%%%%%%%%%%%%%%%%%%%%%%%%%%%%%%%%%%%

\section{Atomistic simulations of the polystyrene melt}
\begin{figure}
\begin{center}
\includegraphics[width=8.0cm,angle=0]{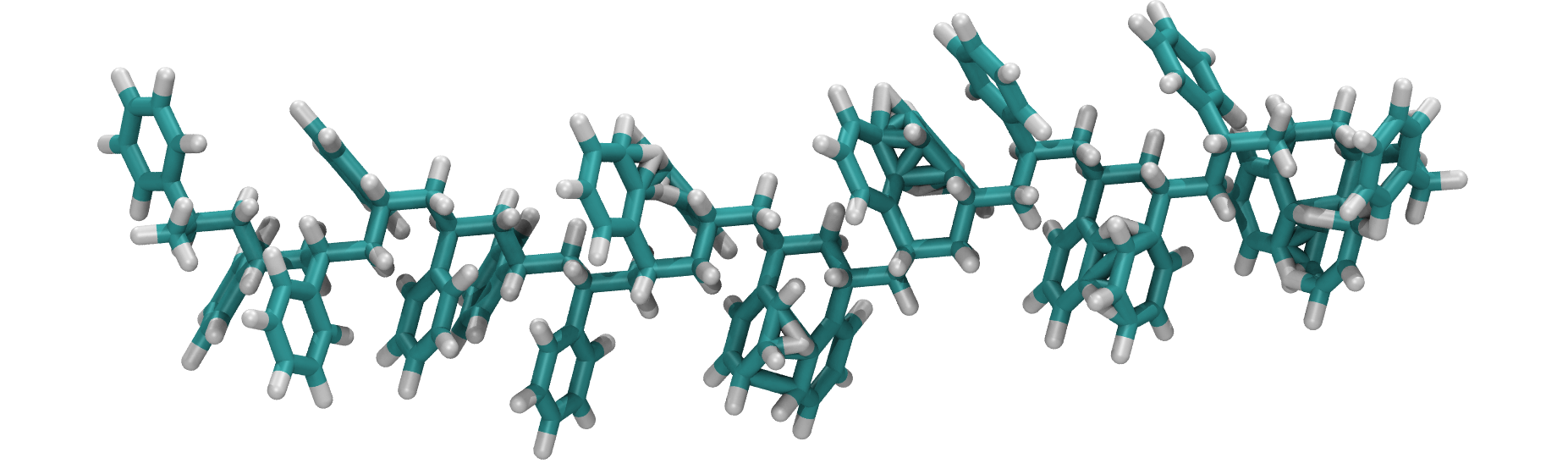}\\
\caption{Atomistic representation of a polystyrene chain where the carbon
atoms are given in blue and hydrogen atoms in white.}\label{fig:PS}
\end{center}
\end{figure}

In the present section we briefly explain how to build a simplified 
model of the
polystyrene melt by using the GROMACS~\cite{Gromacs} MD simulations.
The whole procedure described herein is framed into a didactic approach
and some specific details are missed in order to make the protocol
as clear as possible. 
In this study we have investigated polystyrene chains constituted by
162 atoms, as reported in Fig.~\ref{fig:PS}: specifically, a single
polymer chain is constituted by 80 carbon atoms and 82 hydrogen atoms.
The number of atoms is high enough to model a realistic polystyrene chains, 
but not so high as to become computationally expensive.  
Then, a cubic simulation box containing 68 polystyrene chains (and hence
11016 total atoms) has been built by using the Packmol program~\cite{Packmol},
which allows to randomly put the desired number of atoms and molecules inside a
box with a given volume.
%%%%%%%%%%%%%%%%%%%%%%%%%%%%%%%%%%%%%%%%%%%%%%%%%%%%%%%%%%%%%%%%%
%%%%%%%%%%%%%%%%%%%%%%%%%%%%%%%%%%%%%%%%%%%%%%%%%%%%%%%%%%%%%%%%%
The box length has been fixed to 5 nm, in order to have a volume of 
125 nm$^2$ and therefore a polystyrene density of 950 kg/m$^2$. 
Once built the initial configuration, we have made use of the OPLS 
force field~\cite{Jorgensen:88} to set the intra and intermolecular potentials
between the polystyrene chains. The bond distances between C and H atom,
along with the numerical expression of bond stretching potential $V_b$ and
the force constant $k_r$ are reported in Tab.~\ref{tab:bond}. Here, $r$ and
$r_0$ indicate the bond stretching and the equilibrium atomic distance, 
respectively. 
%%%%%%%%%%%%%%%%%%%%%%%%%%%%%%%%%%%%%%%%%%%%%%%%%%%%%%
\begin{table}
\begin{center}
\caption{Parameters of bond stretching potential
$V_{b}(r)\equiv (k_r/2)(r-r_0)^2$.}
\label{tab:bond}
\begin{tabular*}{0.45\textwidth}{@{\extracolsep{\fill}}cccccc}
\hline
\hline
& Bond & \qquad & $r_0$ (nm) & \qquad & k$_r$(kJ mol$^{-1}$ nm$^{-2})$ \\
\hline
& C-C & \qquad & 0.1380 & \qquad & $10^7$ \\
& C-H & \qquad & 0.1070 & \qquad & $10^7$ \\
\hline
\end{tabular*}
\end{center}
\end{table}
%%%%%%%%%%%%%%%%%%%%%%%%%%%%%%%%%%%%%%%%%%%%%%%%%%%%%%
%%%%%%%%%%%%%%%%%%%%%%%%%%%%%%%%%%%%%%%%%%%%%%%%%%%%%%
\begin{table}
\begin{center}
\caption{Parameters of non-bonded potential
%$V_{nb}(r_{ij})\equiv V_{LJ}(r_{ij})+V_{Coul}(r_{ij})+V_{rf}(r_{ij})$.}\label{tab:pot}
$V_{nb}(r_{ij})\equiv V_{LJ}(r_{ij})$.}\label{tab:pot}
\begin{tabular*}{0.45\textwidth}{@{\extracolsep{\fill}}cccccccc}
\hline
\hline
& Atom & \qquad & $\sigma$ (nm) & \qquad & $\epsilon$(kJ mol$^{-1})$  & \qquad
& $q$(e) \\
\hline
& C & \qquad & 0.375000 & \qquad & 0.439000 & \qquad & 0.000\\
& H & \qquad & 0.235200 & \qquad & 0.092000 & \qquad & 0.000\\
\hline
\end{tabular*}
\end{center}
\end{table}
%%%%%%%%%%%%%%%%%%%%%%%%%%%%%%%%%%%%%%%%%%%%%%%%%%%%%%
In Tab.~\ref{tab:pot} we report all the non-bonded interactions,
i.e. the interactions between atom pairs whose distance $r_{ij}$
is not fixed by the
connectivity. Specifically, in this table we have defined:
\begin{equation}
V_{LJ}(r_{ij}) = 4\epsilon[(\sigma/r_{ij})^{12}-(\sigma/r_{ij})^{6}] \,,
\end{equation}
%
%%%%%%%%%%%%%%%%%%%%%%%%%%%%%%%%%%%%%%%%%%%%%%%%%%%%%%%%%%%%%%%%%
\begin{figure}
\begin{center}
\begin{tabular}{cc}
\includegraphics[width=8.0cm,angle=0]{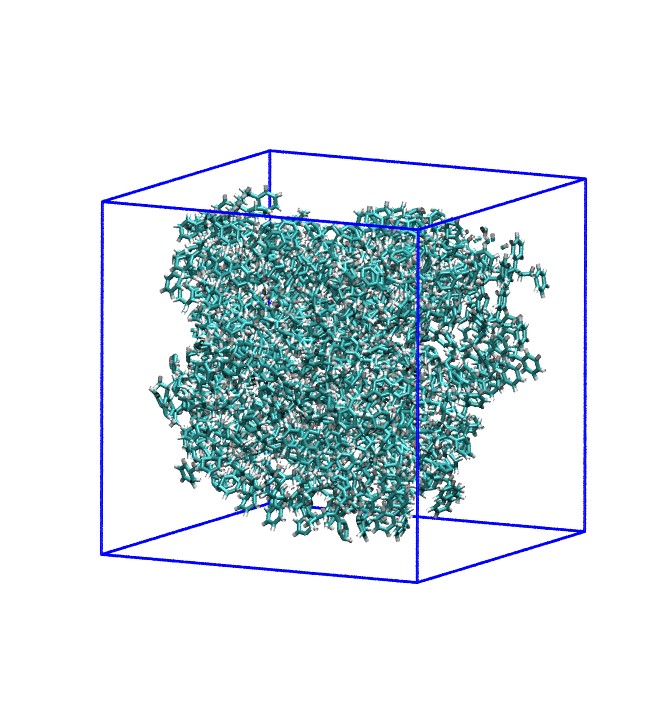}
\includegraphics[width=8.0cm,angle=0]{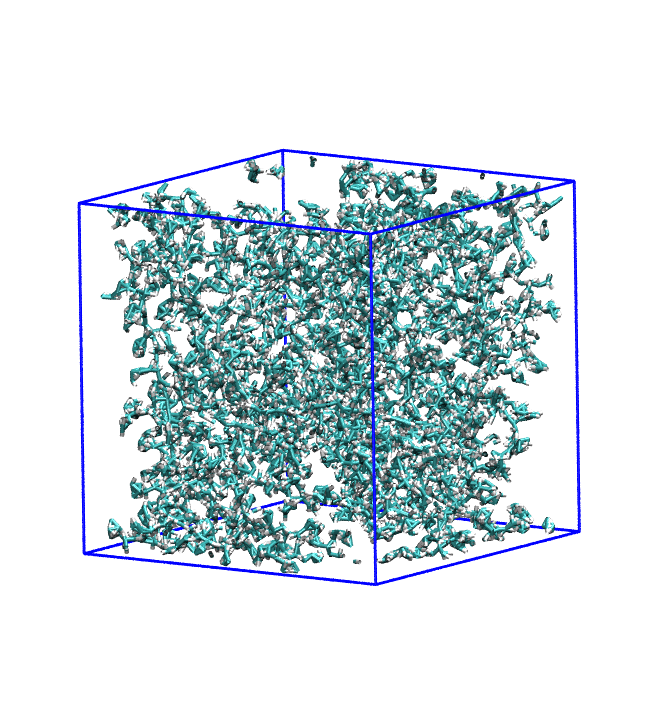}
\end{tabular}
\caption{Simulation snapshots of initial (left) and final (right)
configurations of atomistic polystyrene chains in GROMACS MD 
simulations. The final configuration has been obtained after 80 ps.
The color code is the same as Fig.~\ref{fig:PS}}\label{fig:PS-box}
\end{center}
\end{figure}
%%%%%%%%%%%%%%%%%%%%%%%%%%%%%%%%%%%%%%%%%%%%%%%%%%%%%%%%%%%%%%%%%
%%%%%%%%%%%%%%%%%%%%%%%%%%%%%%%%%%%%%%%%%%%%%%%%%%%%%%%%%%%%%%%%%
\begin{figure}
\begin{center}
\begin{tabular}{cc}
\includegraphics[width=8.0cm,angle=0]{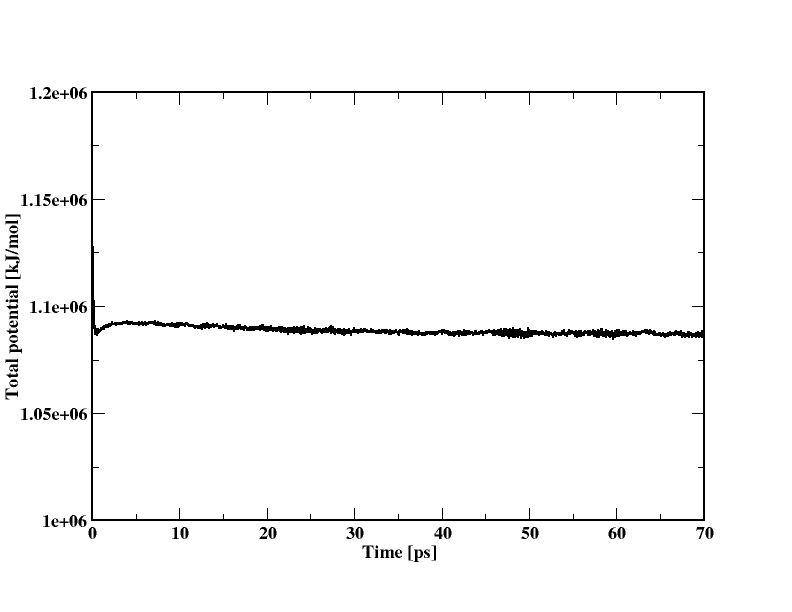}
\includegraphics[width=8.0cm,angle=0]{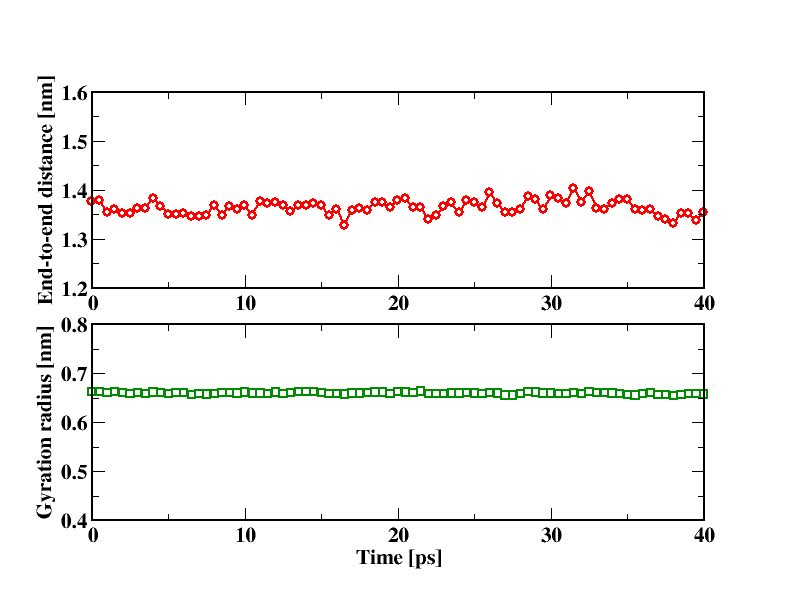}
\end{tabular}
\caption{Left: behavior of the total energy as a function of the simulation
time. Right: end-to-end distance (top) and gyration radius (bottom) of 
polystyrene chains calculated along the production run.}\label{fig:PS-pot}
\end{center}
\end{figure}
%%%%%%%%%%%%%%%%%%%%%%%%%%%%%%%%%%%%%%%%%%%%%%%%%%%%%%%%%%%%%%%%%
corresponding to the Lennard-Jones potential, which is calculated
through the interaction energy $\epsilon$ and the close-contact 
distance $\sigma$. The chosen values for H atoms have been taken from
Ref.~\cite{Munao:18}.
In principle, the total intermolecular potential is given by the sum of
the Lennard-Jones contribution and the Coulombic interaction.  
However, in our simplified approch, the partial charges have been set to zero
everywhere, therefore this contribution does not influence the total 
intermolecular potential. \\
In the present work all atomistic simulations have been performed in a 
cubic simulation box with
periodic boundary conditions. The polymer melt has been simulated in a 
canonical ensemble (NVT), in which temperature, volume and particle number have
been kept fixed. The temperature has been maintained contant at $T= 590 K$
by using the Nose-Hoover thermostat. This temperature has been chosen in order
to guarantee a fast relaxation of the polymer chains. According to the
simulation protocol followed in the present work, we first performed a 
minimization procedure of 10 ps, with a time step of 1 fs, in order to 
remove the possible atoms overlap. After this preliminary run, we have
performed an equilibration run of 30 ps, followed by a production run of 
40 ps, where the final properties of the polymer melts have been calculated.
Two snapshots of the system at the beginning and at the end of the overall 
simulation run are shown in Fig.~\ref{fig:PS-box}. \\
In order to ascertain whether the system reached the thermodynamic 
equilibrium, it is possible to check the behavior of the total system energy
as a function of the simulation time. Indeed, the equilibrium occurs
when the energy does no longer chains with the time. In the left panel 
of Fig.~\ref{fig:PS-pot} the behavior of the system energy as a function of
the simulation time is reported: as visible, after a first decay, the energy
achieves a constant value of $\approx 1.09 \cdot 10^6$ kJ/mol, after 30 ps. 
Therefore, the properties of polystyrene chains have been calculated along the
next 40 ps, corresponding to the production stage. 
Once gained the thermodynamic equilibrium, many different
structural and thermodynamic properties of the system can be calculated.
As an example, we report
in the right panel of Fig.~\ref{fig:PS-pot} the time dependence of the
end-to-end distance and the gyration radius of the polystyrene chains 
investigated in this work. Their average values amount to 1.36 and 0.66 nm,
respectively. More detailed investigations, including the possible addition
of fillers inside the polymer matrix, are outside the scope of the present 
work, but nonethless they can be performed, although more refined approaches
are required. \\
The multiscale approach presented in this work is then completed by 
performing the finite element simulations, whose main results are reported
in the next section.

%%%%%%%%%%%%%%%%%%%%%%%%%%%%%%%%%%%%%%%%%%%%%%%%%%%%%%%%%%%%%%%%%
\section{Finite element simulations}
\begin{figure}
\begin{center}
\includegraphics[width=12.0cm,angle=0]{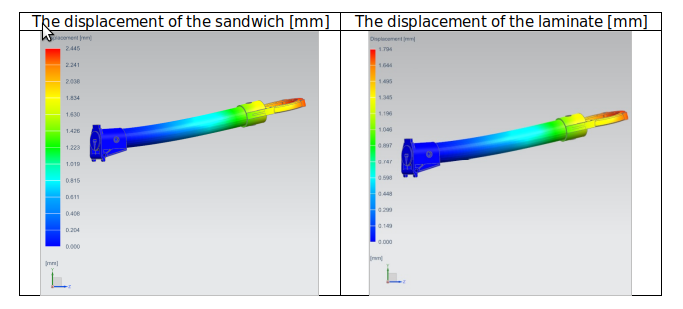} 
\caption{Maximum arm displacement of sandwich (left) and laminate (right).}	
\label{fig:drone10}
\end{center}
\end{figure}

\begin{figure}
\begin{center}
\includegraphics[width=12.0cm,angle=0]{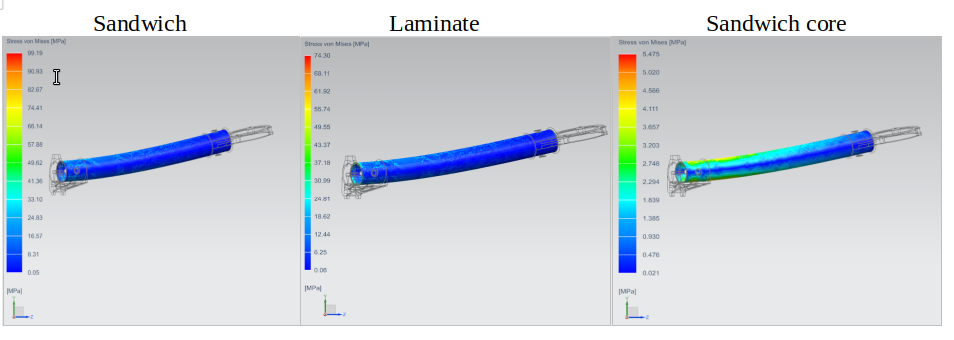} 
\caption{Maximum von Mises stress (in MPa) of sandwich (left), laminate 
(center) and sandwich core (right).}	
\label{fig:drone11}
\end{center}
\end{figure}

\begin{figure}
\begin{center}
\includegraphics[width=10.0cm,angle=0]{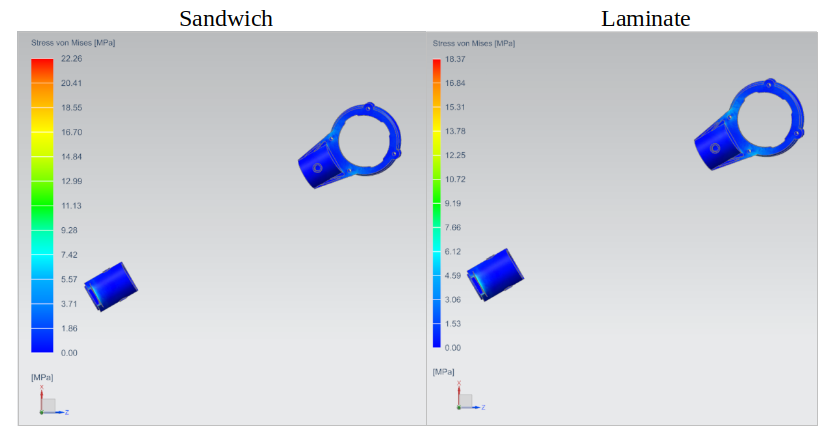} 
\caption{Maximum von Mises stress (in MPa) of aluminum elements of sandwich 
(left) and laminate (right).}	
\label{fig:drone12}
\end{center}
\end{figure}

The finite element simulations have been performed
through the Siemens Nx Nastran 
software and are divided into two steps. The first one is the preprocessing, 
where the mechanical properties of the material are set, including density, 
Young modulus of elasticity and yield stress. In this phase, the mesh is 
also created using the tool inside the software that allows the choice of 
the size of the element and its shape. The smaller the cell size of the 
element, the greater the calculation load of the calculator, and therefore 
the longer the processing time.
The second phase concerns the applications of the boundary conditions, 
such as constraints, forces, contacts between bodies and application of 
frictions. Finally it is possible to choose the type of solution to adopt; 
some possible choices include linear, dynamic, non-linear static 
simulations, as well as the study of the ways of vibrating of a given 
structure. 
In this work we have chosen to use a linear static solution, which, for 
the software at issue, is called SOL 101.
The displacement produced by the applied force is shown in 
Fig.~\ref{fig:drone10}; it is worth noting that the sandwich has lower 
mechanical performances than the laminate used on the commercial drone.
%%%%%%%%%%%%%%%%%%%%%%%%%%%%%%%%%%%%%%%%%%%%%%%%%%%%%%%%%%%%%%%%%
%%%%%%%%%%%%%%%%%%%%%%%%%%%%%%%%%%%%%%%%%%%%%%%%%%%%%%%%%%%%%%%%%
We have then computed the von Mises stress for sandwich and laminate, 
in order to predict the yielding of these materials under complex loading 
from the results of uniaxial tensile tests. Indeed, according to its 
definition, a given material start yielding when the von Mises stress 
reaches the yield stress $\sigma_{\mu}$.
In Fig.~\ref{fig:drone11} it is possible to see that the maximum von Mises 
stress is higher for the sandwich (left panel) in comparison to the 
laminate (center panel). Better 
results would be obtained upon increasing the thickness of the hides at the 
expense of the total weight. The maximum stress occurs at the joint: in 
the right panel of Fig.~\ref{fig:drone11} the most stressed layer and the 
trend of the tensions on the sandwich core are reported. On the latter, it 
may be worth noting the characteristic distribution of the tensions in the 
presence of a bending stress.
%%%%%%%%%%%%%%%%%%%%%%%%%%%%%%%%%%%%%%%%%%%%%%%%%%%%%%%%%%%%%%%%%
%%%%%%%%%%%%%%%%%%%%%%%%%%%%%%%%%%%%%%%%%%%%%%%%%%%%%%%%%%%%%%%%%
In Fig.~\ref{fig:drone12} it is shown the maximum von Mises stress in the 6061 
aluminum elements: we note that this tension is substantially identical as 
well as its distribution in the material in the two cases considered.
%%%%%%%%%%%%%%%%%%%%%%%%%%%%%%%%%%%%%%%%%%%%%%%%%%%%%%%%%%%%%%%%%
%%%%%%%%%%%%%%%%%%%%%%%%%%%%%%%%%%%%%%%%%%%%%%%%%%%%%%%%%%%%%%%%%
Finally, it is worth noting that also the weight of 
the tubular structure have been considered. Indeed, in the preprocessing phase, 
corresponding to the setting of the software of the material, 
the body has been discretized 
with the mesh, and the materials density have been entered.
Once known volume and density of laminate and sandwich, 
the software automatically 
calculates their mass. As a result, the weight of
a laminate in carbon fiber only and epoxy matrix amounts to 28 grams, while 
the weight of a sandwich with a polystyrene core, carbon fiber skins and 
epoxy matrix amounts to 21 grams. 
This result saves 7 grams of weight for each arm of the drone, resulting 
in a total saving of 42 grams. Further improvements could be obtained by
performing a lightening of the drone body also.
In particular, thanks to recent studies on titanium and 
laser processing~\cite{Fotovvati:18} 
it is possible to replace the supporting structure 
with titanium elements printed with additive manufacturing processes.

\section{Conclusions}
In the present work we have proposed a didactic approach suited to 
design the drone structure by making use of finite element and atomistic
simulations, hence giving rise to a multiscale approach. The study is 
motivated by the possibility to equip the drone with proper sensors suited to
measure the concentration of volcanic ash, SO$_2$, CO$_2$ 
and other pollutants in 
the atmosphere. This is particularly interesting in case of volcanic eruptions,
where a massive amount of ash and gases are emitted in a relatively short
time. For such an aim, we have modeled the tubular structure of the drone
with a sandwich constituted by a a polystyrene core,
carbon fiber skins and epoxy matrix. The polystyrene structure has been
also investigated in detail by using atomistic Molecular Dynamics simulations,
which allowed to calculate the end-to-end distance and the gyration radius
of polystyrene chains constituted by 162 atoms. As a main result of the 
present study, we have documented a weight saving of 7 grams for each arm 
of the drone in comparison to the standard commercial drones, although
a slight worsening of the mechanical performances is observed.
The weight saving is limited, but 
the study has focused only to the tubular structure of the drone; 
in future works we plan to show how the weight saving can be improved
by using the topological optimization techniques 
on the body of the drone and on the supports for landing. 

\section*{Acknowledgements}
The present work frames within the PON project titled ``Impiego di tecnologie,
materiali e modelli innovativi in ambito aeronautico AEROMAT'',
avviso1735/Ric, 13 luglio 2017.

\bibliography{manuscript}
\bibliographystyle{ieeetr}
\end{document}